\documentstyle[11pt,newpasp,twoside]{article}
\def\edcomment#1{\iffalse\marginpar{\raggedright\sl#1\/}\else\relax\fi}
\reversemarginpar

\begin{document}
\title{Atomic Data Needs for Modeling Photoionized Plasmas}
 \author{Daniel Wolf Savin}
\affil{Columbia Astrophysics Laboratory, Columbia University, New York,
NY 10027, USA}

\begin{abstract}

Many of the fundamental questions in astrophysics can be addressed
using spectroscopic observations of photoionized cosmic plasmas.
However, the reliability of the inferred astrophysics depends on the
accuracy of the underlying atomic data used to interpret the collected
spectra.  In this paper, we review some of the most glaring atomic data
needs for better understanding photoionized plasmas.

\end{abstract}

\section{Introduction}

Spectroscopic observations of photoionized plasmas can be used to
address many of the fundamental questions in astrophysics.  Planetary
nebulae (PNe) and H II regions provide information about the primordial
He abundance which is an important constraint for theories of
cosmological nucleosynthesis (Pagel 1997).  In addition, PNe can be
used to study the processing and dredging that goes on during the
lifetime of a star (Clegg et al.\ 1987; Boroson et al.\ 1997; Hyung,
Aller, \& Feibelman 1997).  H II regions can also be used to determine
galactic abundance gradients (Pagel 1997).  Stellar winds from hot
stars input energy, momentum, and nuclear processed material into the
interstellar medium and are vital for the evolution of galaxies
(Kudritzki \& Puls 2000).  Active galactic nuclei (AGN) and quasars can
be used to study General Relativity (Nandra et al.\ 1997), the physics
of matter-radiation interaction in extreme environments (Kahn \&
Liedahl 1995), and through measurements of the chemical abundances in
the emitting gas, the chemical evolution of massive galaxies (Hamann \&
Ferland 1993, 1999).  Low-redshift Ly$\alpha$ clouds can be used to
study the origin and chemical evolution of gas at large distances from
galaxies and also to provide constraints on the ionizing spectrum
between 1 and 4 Ryd in order to discriminate between an extragalactic
radiation field due to AGN or due to O stars from starburst galaxies
(Shull et al.\ 1998).  Studies of the IGM yields information on the
chemical evolution of the universe and the shape of the metagalactic
radiation field as a function of redshift (Songaila \& Cowie 1996;
Giroux \& Shull 1997; Savin 2000).

The physical conditions in photoionized plasmas are governed by a host
of microphysical atomic processes.  Ionization of the plasma is due to
photoionization (PI), Auger ionization, collisional ionization, and
charge transfer (CT).  Electron-ion recombination occurs through
radiative recombination (RR), dielectronic recombination (DR),
three-body recombination, and CT.  Line emission is due to RR, DR, CT,
and electron impact excitation (EIE).  Bound-bound transitions due to
photoabsorption can be important.  Free electrons are assumed to have a
predominately Maxwellian velocity distribution with a kinetic
temperature $T_e$ determined by the balance between the processes of
heating (photoelectric, mechanical, cosmic ray, etc.) and cooling
(predominantly inelastic collisions between electrons and other
particles).  Together these various processes determine the ionization
structure, the electron temperature, the level populations of excited
states, and all emission and absorption features.  The observed
spectrum results from transport of radiation through the
depth-dependent physical conditions.  No analytical solutions are
possible because of the intricacies, and large-scale numerical
simulations must be performed instead.

Interpreting the spectra of photoionized cosmic sources is carried out
using spectral simulation codes such as CLOUDY (Ferland et al.\ 1998),
ION (Netzer 1996; Kaspi et al.\ 2001), NEBU (P\'equignot et al.\ 2001)
and XSTAR (Kallman \& Bautista 2001).  Codes which have the
capabilities of modeling stars with extended outflowing atmospheres
(i.e., stellar winds) have been written by Hamann \& Koesterke (2000),
Haser et al.\ (1998), Hillier \& Miller (1998), Short et al.\ (2001),
and others.  Fundamental to the reliability of these simulations and
any inferred astrophysical conclusions is an accurate understanding of
the underlying atomic physics which produces the observed spectra.  In
this overview we will briefly outline some of the most glaring atomic
data needs which were presented at this conference.

\section{Atomic Data Needs}

Reliable atomic data are needed for nearly every atomic process
involving the first 30 elements of the periodic table.  The vast
quantity of needed data is overwhelming.  Given the limited resources
currently devoted worldwide to this task, it will likely take decades
to produce all the needed data.  Here we attempt to identify high
priority atomic data needs.  In specific, our aim is to help
researchers identify and address those uncertainties in our
understanding of atomic physics which have the largest impact on our
ability to address the fundamental questions in astrophysics.

The easiest selection criterion which can be used to prioritize the
needed atomic data is for researchers to focus on the twelve most
abundant elements, namely H, He, C, N, O, Ne, Mg, Si, S, Ar, Ca, and
Fe.  But even within this set of twelve, the amount of needed atomic
data is still vast.  To provide further guidance, we discuss below some
of the specific atomic physics which needs to be better understood for
astrophysics.  Our discussion focuses on those issues which were raised
during the conference as being important for the understanding of
photoionized plasmas.  The review here is complementary to the recent
article by Ferland et al.\ (1998) which also discusses the atomic data
needs for modeling photoionized plasmas.

\subsection{Charge Transfer (CT)}

The importance in photoionized plasmas of CT on H and other atoms, as
well as on H$_2$, has long been recognized (e.g., P\'equignot et
al.\ 1978; P\'equignot \& Aldrovandi 1986).  At the relevant
temperatures ($k_BT_e \sim 1$~eV), CT on H is the dominant
recombination mechanism for many one- to four-times charged ions in
photoionized plasmas (Kingdon \& Ferland 1996).  Surprisingly, modeling
has also shown that CT on H of iron ions five-times charged and higher
can significantly affect photoionized plasmas (Ferland et al.\ 1997).
This may also hold for ions of other elements.

Total CT rates are needed for ionization balance calculations.  State
specific partial rates are needed for line emission predictions.  Most
of the rates currently being used for plasma modeling have been
calculated using a Landau-Zener (LZ) approximation and are not expected
to be reliable to better than a factor of $\sim 3$, but the error may
be even larger.  Recent laboratory measurements have shown LZ
calculations can be off by an order of magnitude (Thompson et
al.\ 2000).  Much theoretical and experimental work is needed to
provide reliable CT rates for modeling cosmic plasmas.

\subsection{Dielectronic Recombination (DR)}

DR rates at the low temperatures appropriate for photoionized plasmas
are theoretically and computationally challenging.  It is particularly
difficult to calculate accurate resonance energies for DR resonances at
electron-ion relative energies close to zero eV.  Laboratory
measurements have consistantly found errors even in state-of-the-art
theory for resonance energies and strengths below $\sim 3$~eV (DeWitt
et al. 1996; Schippers et al. 1998; Schippers et al. 2001; Savin et
al.  2001).  These near zero eV resonances often dominate the DR rate
in photoionized plasmas.

Care also needs to be given to states which are forbidden to autoionize
in LS coupling.  Recent measurements of DR onto C~IV have shown the
importance at low temperatures of DR via these LS forbidden
autoionizing states (Mannervik et al.\  1998; Schippers et al.\ 2001).

Currently, the majority of DR rates for modeling photoionized plasmas
come from older calculations.  These rates, given the computational
power available at the time and the theoretical approximations used to
make the calculations tractable, are expected to be somewhat
unreliable, particularly at low temperatures.  This lack of reliable
low temperature DR rates is one of the dominants uncertainties in the
ionization balance in photoionized plasmas (Ferland et al.\ 1998).

Modern calculational techniques can be used to produce the needed DR
rates.  However, given the discrepancies found between theory and
experiment at low energies, it is clear that improvements to the
theoretical techniques are called for.  Until these needed theoretical
advances are achieved, the only way to produce reliable low temperature
DR rates will involve a co-ordinated effort between laboratory
measurements and theoretical calculations.

\subsection{Electron Impact Excitation (EIE)}

Many plasma diagnostics used for interpreting spectra of AGN, H II
regions, and PNe use EIE-generated line emission from transitions
between low-lying energy levels of various atomic ions (Osterbrock
1989).  The upper state in the observed transitions can be populated by
direct EIE (a one-step processes) or by resonant EIE (a two-step
processes).  Particularly challenging to calculate is the contribution
to the total EIE rate due to resonant EIE.  This channel involves
dielectronic capture to an excited (autoionizing) state of the
recombined system which then autoionizes to the upper level of interest
in the initial ion.  This indirect excitation process is the dominant
population mechanism for many important transitions. For example, for
$^3P_1-\ ^3P_2$ and $^2P_{1/2}-\ ^2P_{3/2}$ transition within the
valence shell, calculations carried out including resonance effects can
yield EIE rates over an order of magnitude larger than calculations
which do not account for resonances (Oliva, Pasquali, \& Reconditi
1996).  For situations where resonance EIE is important, reliable EIE
rates requires knowing the position of those autoionizing levels lying
within $\sim 4$~eV of the ground state (Oliva et al.\ 1996).
Accurately determining these resonance energies is extremely
challenging theoretically and experimentally.

\subsection{Energy Levels}

Reliable energy levels are important for modeling and interpreting
spectra from stellar winds of hot stars.  Bound-bound photoabsorption
transitions between 800 and 1300 \AA\ can dramatically affect stellar
spectrum (Hillier 2001) and type Ia supernovae spectra (Pauldrach et
al.\ 1996).  A significant number of these transitions are due to
states involving electrons in $n \ge 4$ levels, up to $n=15$ in some
situations (Kurucz 2001).  For many ions, no accurate measured
wavelengths exist for the corresponding transitions.  Theoretical
energy levels must be used and the resulting wavelengths can be shifted
by $\sim 1$~\AA\ or sometimes more (Haser et al.\ 1998).  Once the
energy levels are reliably known, then improved (photoabsorption)
oscillator strengths can be calculated.

Of all the cosmically-abundant heavy metals in stellar atmospheres,
iron has the richest spectrum.  Of particular importance for
understanding stellar atmospheres and wind are states involving $n \ge
4$ levels in Fe IV, V, and VI.  The article by Hillier (2001) in this
conference proceedings briefly discusses and demonstrates the
importance of these high-lying levels in models of stellar winds from O
stars.

\subsection{Photoionization (PI)}

The majority of PI cross sections currently used by astrophysicists for
valence shell electrons come from OPACITY Project (OP) calculations
(Opacity Project Team 1995).  The OP energy levels are uncertain by
1-2\% (Verner, Barthel, \& Tytler 1994).  This translated directly into
uncertainties in the resonance structure of the PI cross section and
can introduce significant uncertainties when calculating PI rates due
to spectral lines (Verner et al.\ 1996).  Accurate knowledge of this
resonance structure is particularly important in the wavelength regions
around H I Ly$\alpha$, He I K$\alpha$, and He II Ly$\alpha$, as well as
near other astrophysically important lines.  To reduce the effects of
accidental coincidences between resonances and spectral lines, OP
results are often averaged over resonances (Verner et al.\ 1996,
Kallman \& Bautista 2001).  It is also worth noting that the OP
assumption of $LS$-coupling is not satisfactory for closed shell
systems (Verner et al.\ 1994).

Recent laboratory advances have finally allowed PI measurements of
valence shell for a few astrophysically important singly- and
doubly-charged ions (e.g.\ Kjeldsen et al.\ 1999, 2000, 2001).  Reasonable
agreement has been found between theory and experiment for most of
these systems.  But for Fe II, which is one of the astrophysically most
important ions (Viotti, Vittone, \& Friedjung 1988), significant
discrepancies remain (Kjeldsen 2000).

Innershell PI of ions plays an important role in determining the
ionization and thermal structures and line emission of X-ray
photoionized plasmas.  The innershell PI cross sections currently used
for modeling these plasmas have been calculated using a Hartree-Slater
central field approximation (Reilman \& Manson 1979; Verner et
al.\ 1996).  The calculations do not account for any of the structure
in the PI cross section near the innershell ionization thresholds.  To
date, these calculations have been benchmarked only with measurements
on atoms.

Multielectron ionization can occur as a result of innershell PI.  This
is due to Auger processes which occur as the electronic structure of
the ion relaxes to fill the innershell hole.  Auger yields are
currently estimated using theoretical and experimental results for
atoms from the early 1970's (Kaastra \& Mewe 1993 and references
therein).  These estimates have never been tested for ions and their
reliability is highly questionable.

\section{Conclusions}

Shortcomings in our current theoretical and experimental capabilities
in atomic physics as well as a lack of available atomic data hinder our
ability to infer reliably the properties of many cosmic plasmas.  This
in turn limits our ability to address many of the fundamental issues in
astrophysics.  In order to help improve the situation, we have
attempted to list here some of the most glaring atomic data needs for a
better understanding of photoionized cosmic plasmas.  It is our hope
that this review will help to guide the future research of theoretical
and experimental atomic physicists.  Given the current resources being
devoted to removing the above described uncertainties in our
understanding of atomic physics, it seems likely that it will require
decades to remove these uncertainties and provide data needed by
the astrophysicists.

\acknowledgments

The author thanks Gary Ferland, D.\ John Hillier, Kirk Korista, and Bob
Kurucz for helpful suggestions regarding this article.  DWS is
supported by NASA Space Astrophysics Research and Analysis Program
grant NAG5-5261.

\end{document}